\documentstyle[aps]{revtex}

\begin{document}
\begin{flushright}
{\large\bf To Appear in $<<$Europhysics Letters$>>$}
\end{flushright}
\centerline{\Large {\bf Integrable impurities in Hubbard chain}} 
\centerline{\Large {\bf with the open boundary condition }}

\bigskip
\bigskip

\centerline{\bf Zhan-Ning Hu$^{a,}$\footnote{{\bf E-mail:
huzn@aphy.iphy.ac.cn}} and Fu-Cho Pu$^{a,b,}$\footnote{{\bf E-mail:
pufc@aphy.iphy.ac.cn}}}\centerline{$^a$ Institute of Physics and 
Center for
Condensed Matter Physics,} \centerline{\ Chinese Academy of Sciences,
Beijing 100080, China \footnote{
mail address}} \centerline{$^b$ Department of Physics, Guangzhou 
teacher
colleague, Guangzhou 510400, China}

\begin{center}
\bigskip
\begin{minipage}{5in}




\bigskip
\bigskip

\centerline{\large\bf   Abstract} 
The Kondo problem of two impurities in 1D strongly correlated electron 
system within the framework of the open boundary Hubbard chain is 
solved and the impurities, coupled to the ends of the electron system, 
are introduced by their scattering matrices with electrons so that the 
boundary matrices satisfy the reflecting integrability condition. 
The finite size correction of the ground state energy is obtained 
due to the impurities. Exact expressions for the low temperature 
specific heat contributed by the charge and spin parts of the 
magnetic impurities are derived. The Pauli susceptibility and 
the Kondo temperature are given explicitly. The Kondo 
temperature is inversely proportional to the density of electrons.

\smallskip

\end{minipage}
\end{center}

\newpage

In recent years the Kondo problem\cite{my01} in the low dimensional 
strongly
corrected electron systems has been the focus of intense activity and 
the
low temperature properties of such systems in one spatial dimension 
can be
described in terms of a Luttinger liquid\cite{my0203}. The electron-
electron interaction is discarded in the original treatments of Kondo
problem, which is reasonable in three dimensions where the interacting
electron system is described by Fermi liquid. The recent advances in
semiconductor technology enable to fabricate very narrow quantum wire 
which
can be considered as one-dimensional system and furnishes a real 
Luttinger
liquid. Edge states in a 2D electron gas for fractional quantum Hall 
effect
can also be regarded as Luttinger liquid\cite{my04}. Around these 
subjects
there are intense efforts and much progress has been made from different
approaches by the use of the bosonization and renormalization 
techniques\cite
{my06,my05}, boundary conformal field theory\cite{my08}, the scaling 
method
\cite{my07} and the density matrix renormalization group calculations.

Kondo impurities play an important role in strongly electron models,
especially in one dimensional system, which is due to only a small 
amount of
defects may change the properties of the system, and the impurities 
usually
destroy the integrability of the pure system when they are introduced. 
In
the pioneered work, Andrei and Johannesson\cite{andr} introduced a 
spin-$S$
impurity into the isotropic Heisenberg chain with preserving the
integrability of the system. It was developed in Ref. [9] by Lee and
Schlottmann for the host chains with the higher spin values. Eckle, 
Punnoose
and R\"{o}mer studied some physical properties of the 
model\cite{romer3}.
Recently, Bed\"{u}fig $et$ $al$ has solved an integrable impurity model
where the impurity is coupled to $t-J$ chain with both spin and charge
degrees of freedom\cite{my13}. The impurity is introduced through a 
local
vertices as in Ref. [8]. Schlottmann and Zvyagin introduce the impurity 
in
supersymmetric $t-J$ model via its scattering matrix with the itinerant
electrons\cite{schloTJ} and the Hamiltonian of the system can be 
constructed
by the transfer matrix. They have discussed also the magnetic 
impurities
embedded in the Hubbard model \cite{021sch}. For all these cases the 
$
periodic$ conditions have been imposed on the electron host and the 
impurity
models suffer the lack of backward scattering. Generally speaking, the
impurities can divide a `pure' system into several part with the 
impurities
located at the ends of the every part. And based on Kane and Fisher's
observation \cite{my05}, we see it is advantageous to use open boundary
problem with the impurities at open ends to study the problem of 
impurities
coupled with strongly-correlated electron system. Indeed, this program 
has
been used for $\delta -$ interacting fermi system in Ref. [14] , $t- 
J$
model in Refs. [15-18] and Heisenberg spin chains in Ref.\cite
{ours1,ours2,ours3} with the $open$ boundary conditions.

The one-dimensional Hubbard chain describes the strongly correlated 
electron
system and there are many works since the exact solution was found by 
Lieb
and Wu\cite{lieb} with the periodic condition. The one-dimensional 
Hubbard
chain with open boundary conditions\cite{add01}, of boundary chemical
potentials\cite{add02} and boundary magnetic 
fields\cite{add03,add04}, is
solvable also by the use of the Bethe ansatz method. Recently, E$\beta 
$ler
and Frahm\cite{add05} study the singularities in X-ray absorption 
spectra of
the one-dimensional Hubbard model and the $t-J$ chain and some 
thermodynamic
properties of the model at zero temperature. There are various complex
solutions based on the Bethe ansatz equations and described the 
boundary
bound states for antiholons, spinnons, and pairs of electrons, 
respectively 
\cite{add06}. The finite-size spectrum for the relevant boundary 
conditions
is given in Refs. [29,24,27]. The effect of boundary point contact
potentials on the coherent mesoscopic oscillations is studied by Frahm 
and
Zvyagin\cite{add08} for the strongly correlated quantum wires.

In this letter, we devote to study the magnetic impurities in the $open$
boundary Hubbard chain by the Bethe ansatz technique and to discuss 
the low
temperature properties contributed by impurities with the use of the
Landau-Luttinger description. The quantum inverse scattering 
method\cite
{korebook} is adapted to solve the eigenvalue problem of the transfer 
matrix
which determine the Hamiltonian of the impurity model. As is well-
known, the
scattering matrix for two electrons in the Hubbard chain is 
\begin{equation}
S_{ij}(k_{i,}k_j)=\frac{\sin k_i-\sin k_j-\frac{iU}2P_{ij}}{\sin 
k_i-\sin
k_j-\frac{iU}2}
\end{equation}
where $P_{ij}$ is the permutation operator of the electrons. The wave
function of the chain in region $0\leq x_{Q1}\leq x_{Q2}\leq \cdots 
\leq
x_{QN}\leq G-1$ has the form: 
\[
\Psi _{\sigma _1,\sigma _2,\cdots ,\sigma 
_N}(x_1,x_2,\cdots ,x_N)\quad
\quad \qquad \qquad \qquad \qquad \qquad \qquad \qquad \qquad \qquad 
\qquad
\qquad 
\]
\begin{equation}
=\sum_{\stackrel{P}{r_1\cdots r_N=\pm 1}}\varepsilon _P\varepsilon
_rA_{\sigma _{Q1},\cdots ,\sigma _{QN}}(r_{PQ1}k_{PQ1},\cdots
,r_{PQN}k_{PQN})\exp [i\sum_{j=1}^Nr_{Pj}k_{Pj}x_j]]
\end{equation}
where $\varepsilon _P=1$ $(-1)$ when the parity of $P$ is even (odd)$
,\varepsilon _r=\prod_{j=1}^Nr$ in which $r$ takes the value $+1$or 
$-1,$
which gives that $A_{\cdots \sigma _i,\sigma _j\cdots }\left( \cdots
k_i,k_j\cdots \right) $ $=S_{ij}(k_{i,}k_j)$ $\times A_{\cdots \sigma
_j,\sigma _i\cdots }\left( \cdots k_j,k_i\cdots \right) .$ The 
impurities
are coupled to the two ends of the system and they are introduced by 
the
scattering matrix with electrons, 
\begin{equation}
S_{L,R}\left( k_j,\sigma _j\right) =\frac{\sin k_j-iC_{L,R}-\frac{iU} 
2
P_{L,R\ j}}{\sin k_j-iC_{L,R}-\frac{iU}2},
\end{equation}
so that the boundary matrices take their form as 
\begin{equation}
R_L\left( k_j,\sigma _j\right) =e^{i\varphi 
_L\left( k_j\right) }\frac{\sin
k_j-iC_L-\frac{iU}2P_{L\ j}}{\sin k_j+iC_L+\frac{iU}2P_{L\ j}},
\end{equation}
\begin{equation}
R_R\left( -k_j,\sigma _j\right) =e^{-2ik_j(G+1)}e^{i\varphi _R\left(
k_j\right) }\frac{\sin k_j-iC_R-\frac{iU}2P_{R\ j}}{\sin 
k_j+iC_R+\frac{iU} 2
P_{R\ j}}
\end{equation}
where the interacting parameters $C_{L,R}$ is arbitrary. The boundary 
$R$
matrix satisfies the reflecting Yang-Baxter equation and the 
eigenvalue
problem of the system can be reduced to the form 
\[
\left. Tr_\tau \left[ T\left( \lambda \right) T^{-1}\left( -\lambda 
\right)
\right] \right| _{\lambda =\sin k_j}\Phi \qquad \qquad \qquad \qquad 
\qquad
\qquad 
\]
\begin{eqnarray}
&=&\frac{i\frac U2-\sin k_j}{i\frac U4-\sin k_j}\frac{\sin 
k_j+iC_L+\frac{iU
} 2}{\sin k_j-iC_L-\frac{iU}2}\frac{\sin k_j+iC_R+\frac{iU}2}{\sin 
k_j-iC_R- 
\frac{iU}2} \\
&&\cdot e^{-i\left\{ \varphi _L\left( k_j\right) +\varphi _R\left(
k_j\right) \right\} }e^{2ik_j\left( G+1\right) }\Phi  \nonumber
\end{eqnarray}
with the transfer matrix defined by 
\begin{eqnarray*}
T\left( \lambda \right) =S_{\tau j}\left( \lambda \right) S_{\tau 
0}\left(
\lambda \right) \cdots S_{\tau j-1}\left( \lambda \right) S_{\tau 
j+1}\left(
\lambda \right) \cdots S_{\tau N+1}\left( \lambda \right) ,
\end{eqnarray*}
and $S_{\tau l}\left( \lambda \right) =\left( \lambda -\sin k_l- 
iUP_{\tau
l}/2\right) /\left( \lambda -\sin k_l-iU/2\right) $ with $\sin 
k_0=iC_L$, $
\sin k_{N+1}$ $=$ $iC_R$ for $l=0,1,2,\cdots ,N+1$. Notice that the 
state
function $\Phi $ is related to the coefficient $A_{\sigma _{Q1},\sigma
_{Q2},\cdots ,\sigma _{QN}}(r_{PQ1}k_{PQ1},r_{PQ2}k_{PQ2},\cdots
,r_{PQN}k_{PQN}),$ which is dependent on also the spins of magnetic
impurities and is suppressed for brevity. $P_{j0}$ and $P_{jN+1}$ are 
the
permutation operators between the magnetic impurities and the 
conduction
electrons. The Hamiltonian of the model has the form 
\begin{equation}
H=-\sum_{\left\langle ij\right\rangle ,\sigma =\downarrow \uparrow
}C_{i\sigma }^{+}C_{j\sigma }+U\sum_in_{i\uparrow }n_{i\downarrow }+ 
J_L{\bf 
S }_1\cdot {\bf d}_L+V_Ln_1+J_R{\bf S}_G\cdot {\bf d}_R+V_Rn_G
\end{equation}
where $d_L$ and $d_R$ are the spin operators of the impurities with 
the
spins $1/2$. The interacting constant $J_{L,R}$ and the scattering 
potential 
$V_{L,R}$ of the impurities can be expressed as $
J_{L,R}=4V_{L,R}/3=-2/(1+C_{L,R}^2)$ when $U=2$ and $
J_{L,R}=4UV_{L,R}/(4+U)=U\left( U^2-4C_{L,R}^2-8+\sqrt{
16C_{L,R}^4-8C_{L,R}^2U^2+64C_{L,R}^2+U^4}\right) /\left( 8- 
2U^2\right) $
when $U\neq 2$ where we have assumed that the electrons scatter with 
the
impurities with the small momentums. From the Bethe ansatz equations 
of this
system with the impurities, an immediate consequence is that the total
momentum of the system has the finite size correction $\sum_{j=1}^N
\sum_{l=0,L,R}\theta \left( q_j,2u+C_l\right) /\left( 2G\right) $ ,
contributed by the magnetic impurities coupled to the open boundary 
fermions
system. For the Hubbard ring, $I_j$ and $J_\alpha $ are consecutive 
integers
( or half-odd integers ) centered around the origin and satisfying $
\sum_jk_j=0$ of the ground state. Now the magnetic impurities affect 
the
distributions of $I_j$ and $J_\alpha $ and the index $\alpha $ runs 
from $-M$
to $M$ which is different from the periodic case.

In the thermodynamical limits: $N\rightarrow +\infty ,$ $G\rightarrow
+\infty ,$ $M\rightarrow +\infty $ with the ratios $N/G,$ $M/G$ kept 
finite,
the real numbers $k$ and $\lambda $ are distributed continuously in 
the
ranges $-Q$ to $Q$ $\leq \pi $ and $-B$ to $B\leq +\infty $ with the 
density
functions $\rho \left( k\right) $ and $\sigma \left( \lambda \right) 
$ ,
respectively. The distributional functions satisfy the coupled 
integral
equations 
\begin{equation}
\rho \left( k\right) +\frac{\cos 
k}{2G}\sum_{l=L,R}a\left( q,2u+C_l\right) =
\frac 1{2\pi }+\frac{\cos k}2\int_{-B}^B\sigma \left( \lambda \right) 
\left[
a\left( q-\lambda ,u\right) +a\left( q+\lambda ,u\right) \right] 
d\lambda ,
\label{hu002}
\end{equation}
\begin{equation}
\sigma \left( \lambda \right) +\frac 12\sum_{l=\pm 1}\int_{-
B}^B\sigma
\left( \lambda ^{\prime }\right) a\left( \lambda -l\lambda ^{\prime
},2u\right) d\lambda ^{\prime }=\frac{\sigma ^G\left( \lambda 
\right) }{2G} +
\frac 12\sum_{l=\pm 1}\int_{-Q}^Q\rho \left( k\right) a\left( \lambda
-lk,u\right) dk
\end{equation}
with $\sigma ^G\left( \lambda \right) =a\left( \lambda ,u\right) 
+a\left(
\lambda ,u+C_l\right) +a\left( \lambda ,u-C_l\right) +a\left( \lambda
,u+C_R\right) +a\left( \lambda ,u-C_R\right) $ where $a\left( q,\eta 
\right)
=\pi ^{-1}\eta /\left( q^2+\eta ^2\right) .$ $Q$ and $B$ are the cutoffs 
of $
k$ and $\lambda $ , respectively, for the ground state and they satisfy 
that 
$\int_{-Q}^Q\rho \left( k\right) dk=N/G$ and $\int_{-B}^B\sigma 
\left(
\lambda \right) d\lambda =M/G$ . The ground state energy is $
E/G=-2\int_{-Q}^Q\rho \left( k\right) \cos kdk.$ For the half-filled 
band $
Q=\pi $ and $N/G=1.$ The number of the down spins is a half of that 
for the
conduction electrons in the system. To see this, by Fourior 
transformation
of the second Bethe ansatz equation and assuming that the distributed
functions are the even functions, we have that $\int_{-Q}^Q\rho \left(
k\right) dk=2$ $\int_{-\infty }^\infty \sigma \left( \lambda \right)
d\lambda $ with the choice of $\omega =0$ and $G\rightarrow +\infty 
$. Then
the magnetization is that $S_z=\left( N+2-2M\right) /2=1$, which is
different from the pure system. The density functions are 
\begin{equation}
\widetilde{\sigma }\left( \omega \right) =\frac{J_0(\omega )}{2\cosh 
\left(
u\omega \right) }+\frac 1{2G}\frac{\widetilde{\sigma }^G\left( \omega
\right) }{1+\widetilde{a}\left( \omega ,2u\right) },  \label{hu004}
\end{equation}
\begin{equation}
\rho \left( k\right) =\frac 1{2\pi }+\frac{\cos k}\pi \int_0^\infty 
d\omega 
\frac{\cos \left( \omega q\right) J_0(\omega )}{1+e^{2u\omega }}- 
\frac{\rho
^G\left( k\right) }{2G}+\frac 1{2G}\frac{\cos k}{2\pi }\int_{-\infty
}^\infty d\omega \frac{\widetilde{\sigma }^G\left( \omega \right)
e^{-iq\omega -\left| u\omega \right| }}{1+\widetilde{a}\left( \omega
,2u\right) }  \label{hu005}
\end{equation}
where $\rho ^G\left( k\right) =\cos 
k\sum_{l=L,R}a\left( q,2u+C_l\right) $
and we have used that $u>0$ and $J_0$ is the Bessel function. The terms 
with
factor $1/\left( 2G\right) $ describe the finite size corrections due 
to
impurities. Then the finite size correction for the ground state energy 
due
to the impurities is 
\begin{equation}
E^{\prime }=\int_{-\pi }^\pi dk\cos k\rho ^G\left( k\right) -\int_{- 
\infty
}^\infty d\omega \frac{J_1(\omega )\widetilde{\sigma }^G\left( \omega
\right) }{2\omega \cosh \left( u\omega \right) }
\end{equation}
where $\widetilde{\sigma }^G\left( \omega \right) $ is the Fourior
transformation of the function $\sigma ^G\left( \lambda \right) $. In 
order
to investigate the conductivity properties of the ground state, we 
should
computer the chemical potentials $\mu _{+}$ and $\mu _{-}$ (see Ref. 
[22]).
By the duality of the particles and holes in the system, for the case 
of the
half-filled band, it should be derived that $\Delta =U-2\mu _{-}$ with 
\begin{equation}
\frac{\mu _{-}}G=-4\int_0^\pi dk\cos k\left[ \rho \left( k\right) 
-\rho
\left( k\right) ^{\prime }\right] +\frac 2G  \label{hu003}
\end{equation}
where the density function $\rho \left( k\right) ^{\prime }$ is 
determined
by the coupled integral equations (\ref{hu002}) and 
\[
\int_{-\infty }^\infty d\lambda ^{\prime }\sigma \left( \lambda 
^{\prime
}\right) a\left( \lambda -\lambda ^{\prime },2u\right) +\sigma \left(
\lambda \right) =\int_{-\pi }^\pi dk\rho \left( k\right) 
a\left( \lambda
-q,u\right) +\frac{\sigma ^G\left( \lambda \right) -2a\left( \lambda
,u\right) }{2G}. 
\]
It is due to that a hole with $k_h=\pi $ and $\lambda \rightarrow \infty 
$
appeared comparing with the `half-filled' band and this gives the term 
$
a\left( \lambda ,u\right) /G$ in the above equation and term $2/G$ in
relation (\ref{hu003}). The interesting thing is that the impurities 
do not
change the conductivity property of the system for the ground state 
although
the density functions have been corrected into the forms (\ref{hu004}) 
and ( 
\ref{hu005}).

The other physical properties of the system can be discussed by 
evaluating
its thermodynamics from the Bethe ansatz equations. As an alternative, 
both
more practically and more physically, we now use the picture of the
Landau-Luttinger liquid, putting forward by Carmelo and co- 
workers\cite
{1818} , to determine the magnetic contributions to the specific heat,
susceptibility and Kondo temperature. The densities of the states of 
the
quasiparticles at the Fermi energy $E_F=0$ are described by 
\[
N_c(0)=\frac 1{2\pi v_c}\left[ 1+\frac 1{2G}\frac{\delta \rho \left(
k_0\right) }{\rho \left( k_0\right) }\right] ,\quad N_s(0)=\frac 
1{2\pi v_s}
\left[ 1+\frac 1{2G}\frac{\delta \sigma \left( \infty \right) }{\sigma
\left( \infty \right) }\right] , 
\]
where $v_c$ and $v_s$ are the velocities of the charge and spin
fluctuations, respectively. $\rho $ and $\sigma $ are the distributed
functions in the infinite limit. And $\delta \rho $, $\delta \sigma 
$ are
the finite-size corrections due to impurities. It means that the 
density
functions are denoted by $\rho +\delta \rho /\left( 2G\right) $ and 
$\sigma
+\delta \sigma /\left( 2G\right) $ of parameters $k$ and $\lambda $ ,
respectively. The densities of the states determine the low-
temperature
behaviors of the system by the use of the standard expressions of the
Fermi-liquid theory and we get the contributions to the Kondo effects 
due to
the magnetic impurities as the following form, 
\begin{equation}
\delta C=\frac{\pi \delta \rho \left( k_0\right) }{6Gv_c\rho \left(
k_0\right) }T+\frac{\pi \delta \sigma \left( \infty 
\right) }{6Gv_s\sigma
\left( \infty \right) }T,
\end{equation}
where the specific heat comes from the charges and the spins. The
finite-size correction to the susceptibility is $\delta \chi =\chi 
_0\delta
\sigma \left( \infty \right) /\left[ G\sigma \left( \infty \right) 
\right] $
where $\chi _0$ is the susceptibility in the bulk. The Kondo temperature 
$
T_k $, corresponding to the Fermi temperature in the local Fermi liquid
generated by the impurities, can be derived from the impurity specific 
heat
contributed by the spin sector. It has the form $T_k=2G\sigma 
\left( \infty
\right) /\left[ \pi n\delta \sigma \left( \infty \right) \right] $ 
where $n$
is the density of the electrons in the system. Furthermore, under the 
limits
of the parameters $C_R=C_L=0$, the specific heat due to the spin of 
the
impurity, the susceptibility and the Kondo temperature are 
\[
\delta C_s=\frac{5\pi T}{6Gv_sJ_0\left( i\pi \right) },\delta \chi 
=\frac 5{
GJ_0\left( i\pi \right) }\chi _0,T_k=\frac{2GJ_0\left( i\pi 
\right) }{ 5n\pi 
}. 
\]
In this case we have that $J_{L,R}=4UV_{L,R}/(4+U)=-U$. Notice that 
the
Kondo temperature $T_k$ is linear in the $1/n$ and this property is 
similar
as the one in Ref. [29]. The specific heat contributed by the charge 
of the
impurities have the form: 
\begin{equation}
\delta C_c=\frac{\pi \left[ \ln 2+2\beta \left( 3/2\right) \right] }{
6Gv_cu\left[ 2\sum_{l=1}^\infty \left( -1\right) ^{l-1}\left(
1+4u^2l^2\right) ^{-1/2}-1\right] }T
\end{equation}
with $C_R=C_L=U/4$ and under the other cases the specific heat $\delta 
C_c$
due to impurities have the similar expressions at the low temperature. 
Here
the $\beta $ function is defined by $\beta \left( x\right)
=\sum_{k=0}^\infty \left( -1\right) ^k/\left( x+k\right) $, and $\beta
\left( 3/2\right) \approx 0\allowbreak .\,\allowbreak 429\,2$.

In summary, we have studied the low energy properties of the Kondo 
problem
in a 1D chain of strongly interacting electrons described by the open
boundary Hubbard model. The magnetic impurities are coupled to the ends 
of
the system and they are introduced by the scattering matrices with the
electrons. This is the first approach to study the impurity properties 
in
the $open$ boundary Hubbard chain. The finite size correction of the 
ground
state energy is obtained and we find that the magnetic impurities do 
not
change the conductivity for the ground state although the distributed
functions have been changed. By the use of the Landau-Luttinger liquid
description for the Kondo problem, we obtained the expressions of the
low-temperature specific heat, the susceptibility and the Kondo 
temperature.
The Kondo temperature is inversely proportional to the density of 
electrons.
Finally, we point out that the integrability of the impurity model is
preserved due to the boundary matrices satisfy the reflection equation. 
The
corresponding Hamiltonian has been written down explicitly when the
electrons scatter with the impurities with the small momentums, which 
has a
simple and compact form. The general expression of the Hamiltonian can 
be
constructed with the use of the quantum inverse scattering method and 
we
wish to remain it as an open problem for the further investigations.

\end{document}